# Complete spatiotemporal and polarization characterization of ultrafast vector beams

**Benjamín Alonso,[1,2,*] Ignacio Lopez-Quintas,[1] Warein Holgado,[1] Rokas Drevinskas,[3] Peter G. Kazansky,[3] Carlos Hernández-García[1] and Íñigo J. Sola[1]**

[1]Grupo de Investigación en Aplicaciones del Láser y Fotónica, Departamento de Física Aplicada, Universidad de Salamanca, Salamanca, E-37008, Spain
[2]Sphere Ultrafast Photonics, S.A., R. do Campo Alegre 1021, Edifício FC6, 4169-007 Porto, Portugal
[3]Optoelectronics Research Centre, University of Southampton, Southampton, SO17 1BJ UK
* Corresponding author: b.alonso@usal.es

## ABSTRACT

The use of structured ultrashort pulses with coupled spatiotemporal properties is emerging as a key tool for ultrafast manipulation. Ultrafast vector beams are opening exciting opportunities in different fields such as microscopy, time-resolved imaging, nonlinear optics, particle acceleration or attosecond science. Here, we implement a technique for the full characterization of structured time-dependent polarization light waveforms with spatiotemporal resolution, using a compact twofold spectral interferometer, based on in-line bulk interferometry and fibre-optic coupler assisted interferometry. We measure structured infrared femtosecond vector beams, including radially polarized beams and complex-shaped beams exhibiting both temporal and spatial evolving polarization. Our measurements confirm that light waveforms with polarization evolving at the micrometer and femtosecond scales can be achieved through the use of structured waveplates and polarization gates. This new scale of measurement achieved will open the way to predict, check and optimize applications of structured vector beams at the femtosecond and micrometer scales.

## Introduction

During the last decades, the development of laser technology has boosted our ability to control the properties of ultrafast light pulses. Nowadays it is possible to routinely generate coherent radiation from the near-infrared to the soft x-rays[1], which can be emitted in the form of few-cycle femtosecond laser pulses[2–4], or even attosecond pulses[5]. Furthermore, it is possible not only to tailor their spatiotemporal properties[6–8], but also to structure ultrafast light fields in their angular momentum properties, including both polarization and orbital angular momentum[9–11]. The ultimate control of the angular momentum properties of ultrafast laser pulses has opened new routes for the study of chiral structures[12–14], topological systems[15–17], or magnetic materials[18,19] at the ultrafast timescales.

One example of structured ultrafast light fields with tailored spatiotemporal and angular momentum properties are the so-called vector beams[20]. The paradigm of vector beams is constituted by radially (RP) and azimuthally (AP) polarized beams. In RP and AP beams, the polarization at each point of the beam transverse plane is linear, directed in the radial and tangential directions, respectively. Interestingly, RP beams allow to focus light below the diffraction limit[21], which enables applications in different fields such as laser machining[22,23] or particle acceleration[24,25], among others. AP beams can induce longitudinal magnetic fields at the singularity of the electric field[26] which offers potential applications in spectroscopy and microscopy[27]. Recently, it has been shown that they can be used to decompose produce collinear vortices[28].

Nowadays vector beams can be routinely produced as continuum waves in the infrared (IR) and visible regimes through the use of uniaxial and biaxial crystals[29–31], spatial light modulators[32], optical fibres[33], electrically-tuned q-plates[34] or azimuthally dependent half-waveplates fabricated by ultrafast laser nanostructuring of silica glass, also known as s-waveplates[35], among others. Recently, the generation of short pulsed, femtosecond vector beams[36,37] has gained interest due to their application in high harmonic generation and attosecond science[38–40] or particle acceleration[41]. However, the advances of new laser sources and their applications are tied up to the development of characterization techniques. Since the 1990s, different techniques have been used for the temporal characterization of scalar —i.e., with constant linear polarization (LP)— ultrashort laser pulses[42]. In the last two decades, the problem of measuring spatiotemporal couplings in scalar beams has been tackled with new strategies[43–47]. In parallel, the reconstruction of time-evolving polarization pulses has also been addressed[48–53]. However, the necessity for the full spatiotemporal characterization of structured ultrafast laser pulses, which simultaneously includes both spatiotemporal and polarization properties, remained a challenge up to now.

In this work we implement a technique to characterize the arbitrarily space-time (and the space-frequency) polarization dependence of structured ultrafast fully polarized light pulses. In particular we perform the full characterization of infrared femtosecond vector beams generated through an s-waveplate, which allow us to: (i) to measure the spatiotemporal quality of RP pulses; (ii) to shape and characterize time-dependent vector beams, structured through the use of polarization gates; and (iii) to monitor the focusing properties of structured vector beams. Our technique is based on twofold spectral interferometry, both for the spatiotemporal reconstruction through spatially-resolved spectral interferometry assisted by a fibre-optic coupler[44] and for the polarization analysis through in-line bulk interferometry with a thick birefringent plate[53]. The complete knowledge of the vector beam amplitude and phase allows to reconstruct the polarization state of the beam profile (including intensity, polarization azimuth, and relative phase between the polarization components) both in the space-frequency and space-time domains. We demonstrate that spectral interferometry is a powerful technique for the characterization of ultrafast vector beams, opening the route for a new set of characterization techniques of structured light waveforms, to be used in a diverse range of applications[50,54,55] that make use of spatial and temporal polarization shaping.

**Results**

**Technique for full spatiotemporal polarization measurement**

The ultrafast vector beam can be either expressed in the space-frequency domain or in the space-time domain, being the representation equivalent as they are connected by the Fourier-transform. In the paraxial approximation, the vector beam propagating in the $z$-axis will be characterized by the polarization components in the $xy$ plane, this is to say, $\mathbf{E} = (E_x, E_y)$. In the spatiospectral domain (depending on the angular frequency $\omega$ and the transverse coordinates $x$ and $y$), the electric field components at a certain propagation distance $z$ (observation plane) can be expressed as a function of their amplitudes ($A_x$ and $A_y$) and phases ($\phi_x$ and $\phi_y$), as given by

$$E_x(\omega; x, y) = A_x(\omega; x, y)\exp\left(i\phi_x(\omega; x, y)\right)$$
$$E_y(\omega; x, y) = A_y(\omega; x, y)\exp\left(i\phi_y(\omega; x, y)\right) \tag{1}$$

As we will discuss later, the amplitudes can be directly obtained from the corresponding measured spectra $S$ of the beam, i.e., $A = \sqrt{S}$, while the measurement of the phases conforms the core of the technique.

The technique that we have used to perform the complete spatiotemporal and polarization characterization is based on twofold spectral interferometry. A scheme of the experimental setup is shown in Fig. 1. The laser output is divided into two replicas, one of them is used as a reference (known), while the other beam is shaped in its polarization components –using an s-waveplate optionally with retarder waveplates (in the case of the multiple-order plate, it introduces a delay $\tau_{PG}$ between the fast and slow axes of the plate, due to the difference of dispersion as a result of the birefringence)–, conforming the unknown beam to be characterized. In the unknown beam arm, a thick birefringent plate is placed to introduce a delay $\tau_{YX}$ between the horizontal and vertical polarization components of the beam. Afterwards, a linear polarizer is used to sample different polarization projections of the beam (0°, 45°, and 90° with respect to the $x$-axis polarization direction). While the reference beam is collected by a fixed position fibre port (therefore not being scanned), the unknown beam is spatially scanned (in the transverse $xy$ plane) with the optical fibre port in a motorized 2D translation stage. Notice that it is needed to spatially scan the 3 projections (0°, 45°, and 90°), as well as to scan the $x$-component interference with the fixed reference beam). In the present work, depending on the symmetry of the beam, we have performed in a general case the said scan forming a 2D grid (see Supplementary Note 7) or, when the polarization state does not depend on the radial coordinate, just varying the azimuthal coordinate $\theta$ (i.e., describing a circumference as shown in Supplementary Figure 3). Both single-mode fibres are combined in a broadband fibre-optic coupler. The delay between the reference and the unknown beam, $\tau_{XR}$, is adjusted and fixed with the longitudinal position of the reference fibre. The known reference phase is measured with a standard temporal characterization technique[42] (e.g., in the present case, the spectral phase interferometry for direct electric-field reconstruction (SPIDER) technique[56]).

In order to perform the complete characterization, first we measure the spatially-resolved spectral interferometry between the 0°-projection ($x$-component) of the unknown beam and the reference pulse, $S_{x+Ref}(\omega, x, y)$, by setting the linear polarizer horizontal. Then we obtain their relative phase using Fourier analysis[57] (see details in Methods). Afterwards, we measure the spectrum profile of the same beam projection, $S_x(\omega, x, y)$. As the reference phase is known, we obtain the spatiospectral (and spatiotemporal) amplitude and phase of the $x$-component of the unknown beam. This strategy[44] has been shown to be very versatile by use in the measurement of diffractive focusing, nonlinear processes and few-cycle pulse characterization[8,58–60], among others.

Second, we acquire the 90°-projection of the unknown beam, $S_y(\omega; x, y)$ by placing the linear polarizer vertically. Finally, with the linear polarizer at 45°, we measure an intermediate projection of the $x$ and $y$ components, $S_{x+y}(\omega; x, y)$, which encodes their relative phase (see details in Methods). This allows us to accurately retrieve the phase of the $y$ component, and thus, the frequency-dependent and time-dependent polarization[53]. The phase introduced by the birefringent plate is calibrated as described in the Methods Section. Since the fibre scans the transverse profile of the unknown beam, we retrieve the full spatiotemporal (and spatiospectral) polarization dependence of the beam. The measured polarization ellipses are fully characterized in terms of their ellipticity $\varepsilon = b/a$ and polarization azimuth angle $\chi$ (see inset polarization ellipse in Fig. 1), together with the measured dephase $\delta = \phi_y - \phi_x$ that is related to the ellipticity and provides the handedness of the polarization (positive and negative values of the dephase $\delta$ stand for right- and left-handed polarization, respectively). We redefine the ellipticity as $\varepsilon = \text{sign}(\delta) \cdot b/a$ so it provides the ellipticity and the handedness in a single parameter. Detailed explanation of polarization states and the calculation of $\varepsilon$, $\chi$ can be found in the Supplementary Note 1.

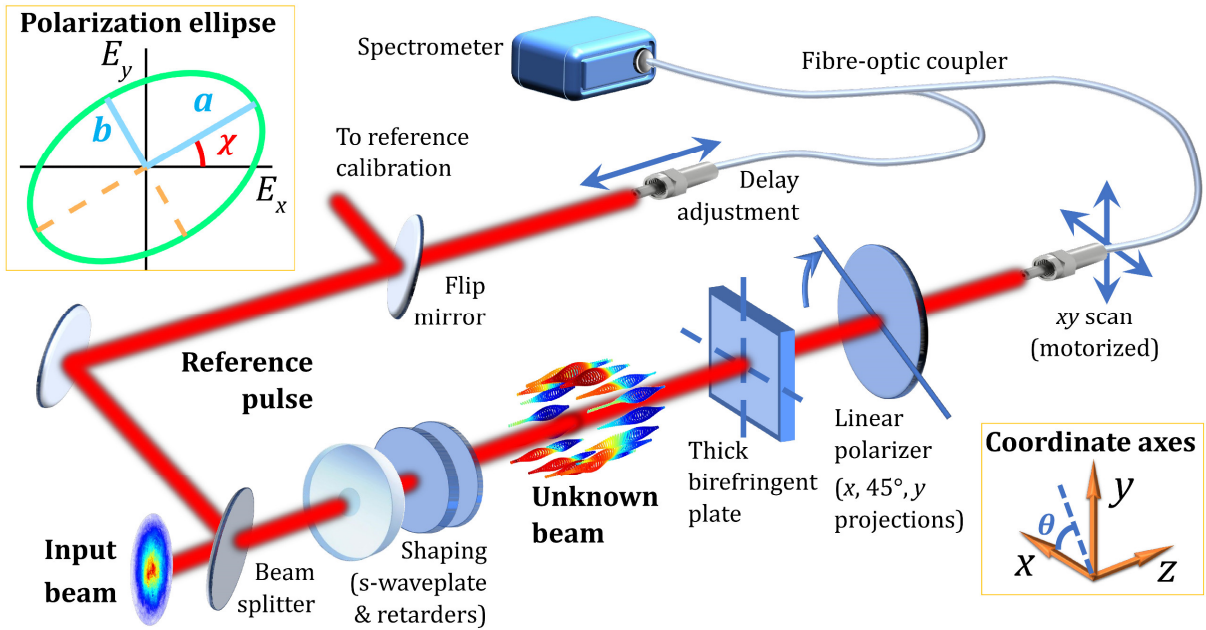

**Figure 1 | Experimental configuration for beam shaping and measurement**. An input beam (in our case, a Gaussian pulsed beam presenting horizontal linear polarization and a time duration of 100 fs at full width half maximum, FWHM) is divided into two arms by a beam splitter. The reference pulse is calibrated by a standard method and collected by one of the two single mode optical fibre port. This port can be moved in the propagation direction in order to adjust a certain delay $\tau_{\text{XR}}$ between the arms of the reference and the unknown beam ($x$ projection), needed for the implementation of the spatio-temporal measurements. The beam in the other arm (unknown beam arm) is shaped by a beam shaper made of an s-waveplate, a multiple-order quarter-waveplate introducing a certain delay $\tau_{\text{PG}}$ between the horizontal and vertical polarization components, and a zero-order quarter waveplate (the last two to create the polarization gate). The $x$ and $y$ components of the polarization shaped (unknown) beam

are delayed with a birefringent plate, introducing a delay $\tau_{YX}$ for the polarization resolved measurements. The unknown beam is spatially scanned with an optical fibre port, that is recombined with the reference beam by means of a fibre coupler. The $x$ projection spectrally interferes with a known reference pulse, while the $y$ and $x$ components spectrally interfere after a 45° linear polarizer. Inset polarization ellipse: scheme of the polarization ellipse defined through the polarization azimuth angle $\chi$ and the ellipticity $\varepsilon = b/a$. Inset coordinate axes: the unknown beam propagates in the $z$-axis, the $x$- and $y$-axes are defined in the transverse plane, where the azimuthal coordinate $\theta$ is measured with respect to the $x$-axis.

## Characterization of radially polarized laser pulses

First, we have characterized a femtosecond RP beam created through an s-waveplate placed after the output of a chirped pulse amplification Ti:sapphire laser (see Methods for further details). In the present case, the radial $r$ and the azimuthal coordinate $\theta$ in the $xy$ plane are uncoupled in the detection plane, provided that the dependence on $r$ is related to the transverse profile of the input beam, which can be considered Gaussian. In order to explore the polarization distribution of the beam along the azimuthal coordinate $\theta$, we scanned the $xy$ plane through a circumference of radius $R = 3$ mm (corresponding to the half-maximum of the beam intensity, see Supplementary Figure 3) around the optical axis. In Figure 2 we show the ellipticity ($\varepsilon$), polarization azimuth ($\chi$) and intensity profiles ($x$-projection, and total) of the vector beam in the space-frequency (first row) and space-time (second row) domains. The $x$-projection of the spectrum or intensity (Figs. 2a and 2b) describes the expected lobes found when a RP bean passes through a linear polarizer horizontally oriented. The $y$-component is complementary so that the total spectrum and intensity (Figs. 2c and 2d) are almost constant across the circular scan. This spatial dependence is inherent to the spatial profile of our laser beam (as shown in Supplementary Note 3). Whereas the ellipticity $\varepsilon$ is found to be close to zero (Figs. 2e and 2f), the polarization azimuth $\chi$—which reflects the orientation angle of the LP field— describes the RP behaviour of the beam (Figs. 2g and 2h), corresponding to the $\chi$-varying LP beam created by the s-waveplate (i.e., the polarization at each point of the beam transversal plane is linear, directed in the radial direction). We find that the operation of the s-waveplate to imprint a RP profile can be considered homogeneous both in the spectral and temporal domains, despite the polychromatic nature of the pulse (notice that the whole spectral bandwidth extends over 25 nm). The full results are shown in Supplementary Note 4. Although in this work we opted to use a radially polarized beam, equivalent results (shifted 90° in the azimuthal coordinate) would be found if using an azimuthally polarized beam (Supplementary Note 5).

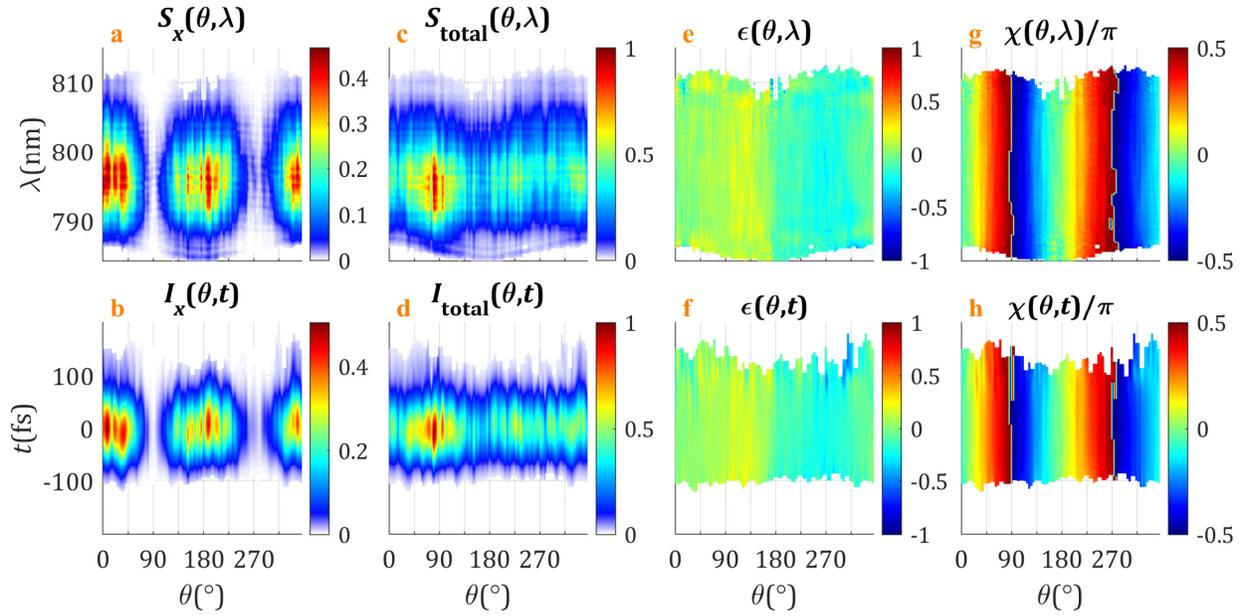

**Figure 2 | Measurement of the beam with radial polarization. a,b** $x$-projection of the spectrum $S_x$/intensity $I_x$; **c,d** total spectrum $S_{\text{total}}$/intensity $I_{\text{total}}$; **e,f** ellipticity $\varepsilon$; and **g,h** polarization azimuth $\chi$. **a,c,e,g:** spatiospectral dependence; **b,d,f,h:** spatiotemporal dependence. Results in the spectral domain are represented as a function of the wavelength $\lambda$, and the time is denoted by $t$. The azimuthal coordinate $\theta$ was sampled in 90 points with a radius of 3 mm. The input laser pulses had a 10 nm (full width at half maximum) spectral bandwidth and 100 fs duration. The results presented in this and following figures are cut below the 1% of the spectral/temporal peak signal. The total spectrum and intensity are normalized to 1.

## Measuring shaped time-dependent ultrafast vector beams

In order to show the ability of our technique to characterize vector beams whose polarization distribution varies temporally in the femtosecond timescale, we structure a laser pulse through the use of polarization gates[61]. When combining an s-waveplate with different types and sets of waveplates, the resulting beam presents a strong spatiospectral and spatiotemporal coupling in its intensity and polarization parameters. One set of waveplates that is particularly interesting is the so-called narrow polarization gate, which is used to effectively generate shorter pulses in certain applications as e.g. isolated attosecond pulse generation[61–63]. In the Methods Section we detail the operation of the narrow gate using a multiple-order quarter-waveplate (QWM) and a zero-order quarter-waveplate (QW0).

In this experiment, we used the RP beam described in the previous Section (created with the s-waveplate), to illuminate the narrow gate elements (i.e. the two waveplates, first the QWM with fast axes at 0° and second the QW0 at 45°) (the narrow gate setup is depicted in the Methods Section). The global effect of the narrow gate is that of producing two circular polarization (CP) pulses delayed 100 fs (similar to the pulse duration) with opposite helicity. When using a RP vector beam, the constant orientation angle (polarization azimuth $\chi$) of the LP light entering the narrow gate depends on the azimuthal coordinate $\theta$, and thus the relative amplitude of both CP pulses depends on $\theta$. When both amplitudes are equal, this is for $\theta = \pi/4 + m\pi/2$ (being $m$ an integer), the narrow gate operates normally (see in Methods), producing LP in the centre and CP in the edges of the pulse in the time domain (Fig. 3a and 3b), whereas the spectral domain ellipticity (Fig. 3c and 3d) is close to $\varepsilon \approx 0$ (as measured in[53]). For the azimuthal coordinates $\theta$ in the RP beam where the linear polarization is parallel to the axes of the QWM (see example in Methods), i.e. $\theta = m\pi/2$, there is only one of the two CP pulses, either being the first (fast axis) or second (slow axis) pulse, which is seen in the temporal intensity (Fig. 3e and 3f), while both, the temporal and spectral ellipticities are $|\varepsilon| = 1$ (constant CP) with the corresponding handedness (left- and right-handed for the pre- and post-pulse respectively), as shown in Fig. 3a-3d. For intermediate azimuthal coordinates $\theta$, the temporal ellipticity minimum ($\varepsilon = 0$) is shifted towards the less intense pulse, being the position where the two overlapping pulses have the same amplitude (Fig. 3a and 3b). In the spectral domain, the x-projection of the spectrum is modulated presenting minima/maxima for $\theta = \pi/4 + m\pi/2$ for the wavelengths where the QWM introduces a dephase $\pm\pi/2$ (exact quarter-wave operation), so its combination with the QW0 results in a vertical/horizontal LP (Fig. 3g and 3h). Therefore, the whole evolution of the polarization shows a strong dependence in the space and time coordinates as seen in the polarization ellipses (Fig. 3i and 3j, and Supplementary Note 6). The effect of the combination of radial polarization and the polarization gate is also manifested in the spatiospectral dependence of the polarization azimuth (Fig. 3k and 3l). The comparison of the experimental results and our numerical simulations performed as described in the Methods Section is shown in Fig. 3. The good agreement between the experimental and theoretical results serves as a validation of our characterization technique. As found in the previous Section, here the small deviations of the experimental results are also due to the spatial profile of the laser beam.

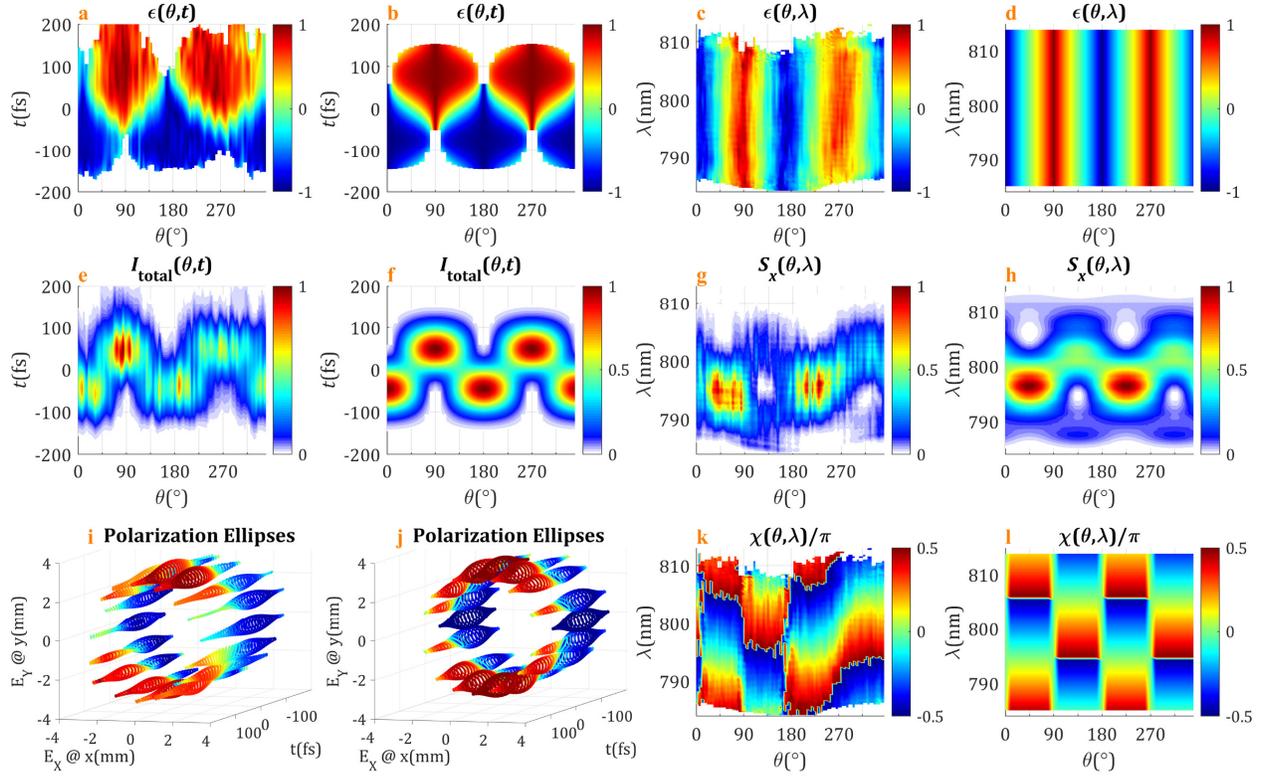

**Figure 3 | Radially polarized beam followed by a narrow polarization gate**. **a,b** spatiotemporal ellipticity $\varepsilon$, **c,d** spatiospectral ellipticity $\varepsilon$, **e,f** spatiotemporal total intensity $I_{\text{total}}$, **g,h** $x$-projection of the spatially-resolved spectrum $S_x$, **i,j** selection of spatiotemporal polarization ellipses ($E_x$ and $E_y$ are the $x$ and $y$ projections of the electric field) for different scanning positions coloured by their time-dependent ellipticity, and **k,l** spatiospectral polarization azimuth $\chi$. **a,c,e,g,i,k**: experiment, **b,d,f,h,j,l**: simulation. The wavelength and the time are denoted by $\lambda$ and $t$, respectively. The azimuthal coordinate $\theta$ is scanned in 90 points with a radius $R$=3 mm. In the ellipticity, the colour scale varies from left-handed circular polarization ($\varepsilon = -1$, blue), to linear polarization ($\varepsilon = 0$, green) and to right-handed circular polarization ($\varepsilon = +1$, red). The $x$ projection of the spectrum and the total intensity are normalized to 1.

**Focusing monitorization of time-dependent vector beam pulses**

Many applications of vector beams, e.g. particle trapping, microprocessing, particle acceleration or nonlinear optics, are carried out at their focus position. Our characterization technique allows us to perform the full characterization both at the far-field and at the near-field. Here we analyse and compare the spatiotemporal polarization dynamics at the focus of the two previously presented vector beams, i.e., the RP vector beam (Fig. 4a-4e) and the RP beam followed by the narrow polarization gate (Fig. 4f-4t). In both cases, the beam was focused using an achromatic lens with a focal length of 50 cm. In these experiments the ultrafast vector beam presents a rich evolution both in the radial and azimuthal coordinates ($r$ and $\theta$). Therefore, we did a two-dimensional spatial scan along the transverse $xy$ plane.

First, when using a purely RP beam, the ellipticity $\varepsilon(x, y, t)$ is ideally zero for every position $x$,$y$ and time $t$. As shown before, although there is a spatiotemporal intensity modulation, there is no substantial temporal dependence of the polarization parameters of the beam. In Fig. 4c, we show the spatial profile at the focus position for a temporal instant that corresponds to the peak of the pulse. We notice that, instead of an ideally homogeneous ring shape of a RP, the measured beam presents spatial intensity modulations due to the above-mentioned inhomogeneities of the input spatial profile. On the other hand, the spatial distributions of the $x$ and $y$ polarization projections (Figs. 4a and 4b, respectively) correspond to that of a RP with the singularity at the centre (see the polarization azimuth $\chi$ profile in Fig. 4e). In the Supplementary Movie 1 we show that the structure of these magnitudes of the focused RP beam is preserved in time.

Contrarily to this case, when focusing the RP beam after passing through the narrow gate, the beam exhibits a temporal/spectral polarization evolution together with the spatial dependence. In this second case the intensity ring is split into two lobes with time-dependent orientations, except for the centre of the pulse at $t$=20 fs (the mean propagating time of the beam components passing through the fast and slow axes of the QWM), where the ring is recovered (Fig. 4m), as in the case of a focused RP beam (Fig. 4c). The $x$ and $y$ projections of this temporal snapshot of the intensity corresponds to two lobes oriented at ±45° (Fig. 4k and 4l). In the temporal leading edge of the beam –corresponding to the fast axis component at QWM– the two lobes are oriented in the $x$-axis for the total intensity as well as for the $x$ and $y$ projections (Fig. 4f, 4g and 4h). On the opposite, in the temporal trailing edge, the two lobes are oriented in the y-axis (Fig. 4p, 4q and 4r), as the slow axis of the QWM is oriented vertically. The complete temporal evolution is shown in Supplementary Movie 2, where the total intensity evolves from two vertical lobes until they completely fill the ring and then they split into two horizontal lobes. Thus, the $x$ and $y$ intensity projections consist in two spatial lobes evolving from vertical to horizontal but rotating in opposite direction. Regarding the ellipticity (Fig. 4i, 4n and 4s), the beam evolves from two dominating left-handed CP horizontal lobes (pre-pulse) to two dominating right-handed CP vertical lobes (post-pulse), with a gradual growing of the latter in detriment of the initial lobes. In between the opposite handedness CP lobes there is LP with a gradual transition. For the temporal maximum (where the total intensity forms a ring), the polarization azimuth $\chi$ (Fig. 4o) of the LP contributions mentioned before is oriented at 0° (horizontal) at the azimuthal coordinate $\theta = 45°$ and $\theta = 225°$ (notice that for those positions the signal is zero for the $y$-projection of the intensity in Fig. 4l), and the LP oriented at 90° (vertical) at the azimuthal coordinate $\theta = 135°$ and $\theta = 315°$ (where the $x$-projection of the intensity in Fig. 4k is zero). The polarization azimuth $\chi$ also evolves gradually from the pre-pulse (Fig. 4j) to the post-pulse (Fig. 4t), alternating the orientation of the LP regions described in Fig. 4i, 4n and 4s (see Supplementary Movie 2).

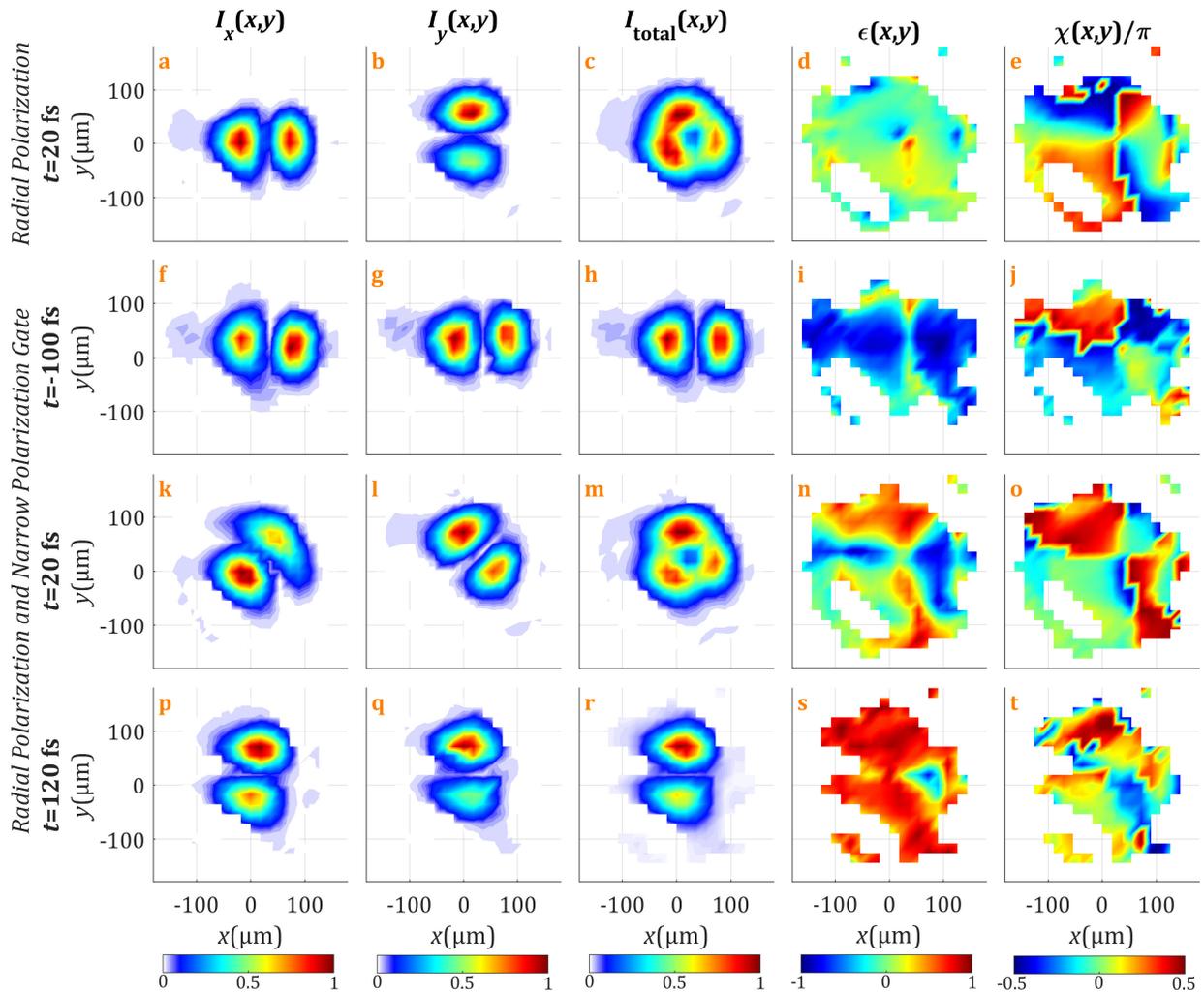

**Figure 4 | Temporal snapshots of the measured focused vector beams**. **a-e**: comparison with the focus for the radially polarized beam at the pulse centre (at time $t$=20 fs). **f-t**: selections for three different times, respectively, **f-j** -100, **k-o** 20 and **p-t** 120 fs, for the radially polarized beam followed by a narrow gate. **a,f,k,p** Horizontal projection of the intensity $I_x$, **b,g,l,q** vertical projection of the intensity $I_y$, **c,h,m,r** total intensity $I_{total}$, **d,i,n,s** ellipticity $\varepsilon$, **e,j,o,t** polarization azimuth $\chi$. The spatial scan was performed in a square grid with 21 x 21 points using a step of 18 µm. All individual intensity maps are normalized to 1.

In the spatiospectral domain, the frequency dependence presents some similarities related to the time domain evolution. First, for the focused radially polarized beam, as the behaviour is achromatic, the results are constant in the spectral domain (as shown in Fig. 5a-5e for 800 nm, and in the Supplementary Movie 3). For the radially polarized beam followed by a narrow gate (Fig.

5f-5t), for some wavelengths (e.g. 795 nm) the $x$ and $y$ projections of the spatially resolved spectrum corresponds to two lobes with opposite orientations at ±45° in the $xy$-plane (Fig. 5f and 5g). Observing the spectral evolution (Supplementary Movie 4), for some wavelengths (e.g. 800 nm) the spatial profile of the $x$ and $y$ projections forms a ring (Fig. 5k and 5l), while for other wavelengths (e.g. 803 nm) the two lobes are crossed at ∓45° with respect to the 795 nm case (Fig. 5p and 5q). The total spectrum forms a ring for every wavelength (Fig. 5h, 5m and 5r). Also, the ellipticity pattern is frequency-independent (Fig. 5i, 5n and 5s), with two horizontal left-handed CP lobes and two vertical right-handed CP lobes and a gradual transition in between (LP at ±45° axes in the $xy$ plane), similarly as described for Fig. 4n (temporal centre of the beam). As expected from the frequency-dependent $x$- and $y$-projections of the spectrum, the polarization azimuth also varies with the wavelength (Fig. 5j, 5o and 5t), being, e.g. for 795 nm (Fig. 5j), $\chi = 0$ in the +45° axis of the $xy$ plane and $\chi = \pm\pi/2$ in the −45° axis of the $xy$ plane. For 803 nm (Fig. 5t) the polarization azimuth is oriented opposite to the 795 nm case.

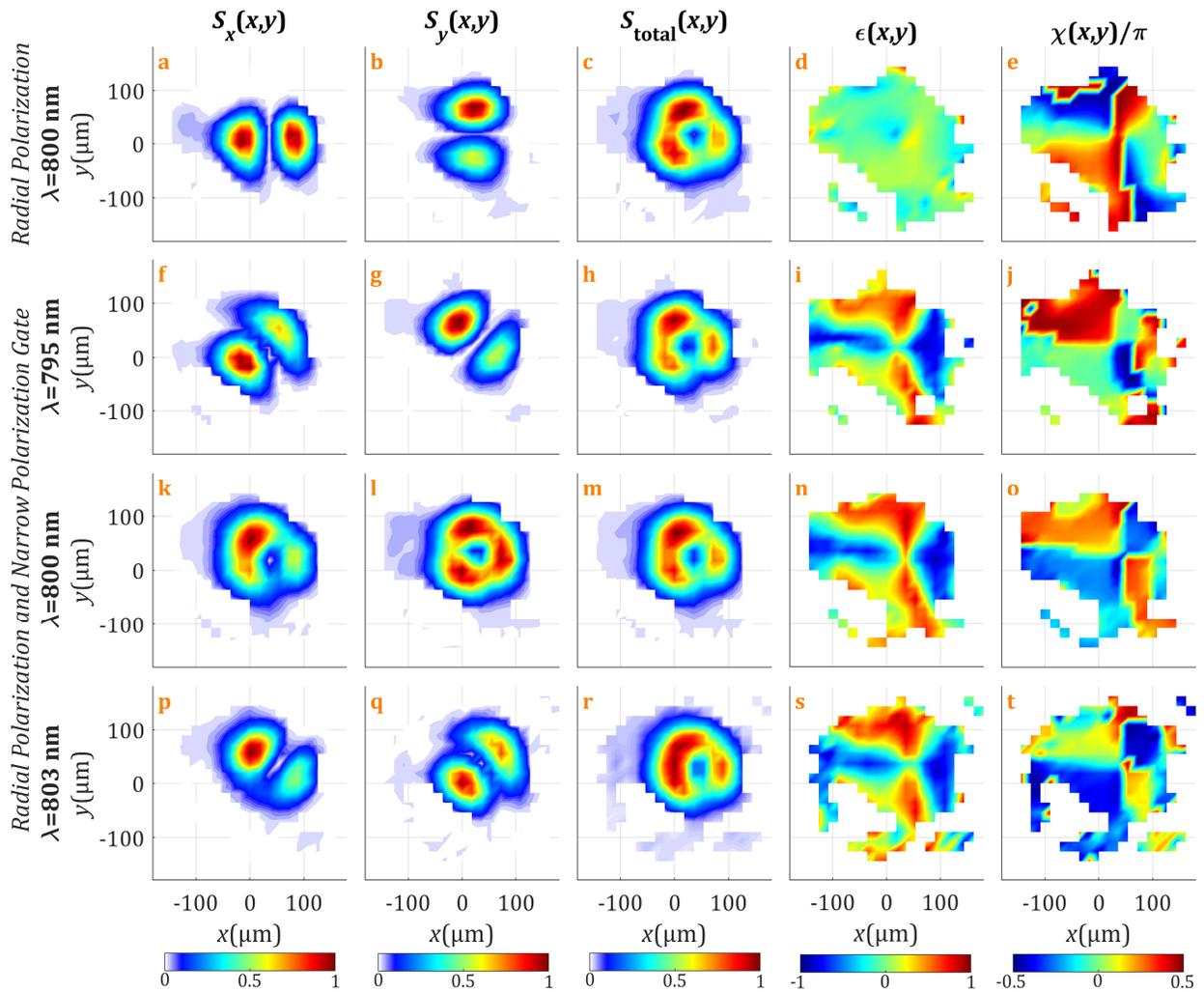

**Figure 5 | Spectral snapshots of the measured focused vector beams**. **a-e**: comparison with the focus for the radially polarized beam at the wavelength λ=800 nm. **f-t**: selections for three different wavelengths, respectively, **f-j** 795, **k-o** 800 and **p-t** 803 nm, for the radially polarized beam followed by a narrow gate. **a,f,k,p** Horizontal projection of the spectrum $S_x$, **b,g,l,q** vertical projection of the spectrum $S_y$, **c,h,m,r** total spectrum $S_{total}$, **d,i,n,s** ellipticity $\varepsilon$, **e,j,o,t** polarization azimuth $\chi$. All individual spectrum maps are normalized to 1.

## Discussion

Tailoring light beams in full dimensionality, i.e., both spatial and temporal shaping of the individual light waveforms on a femtosecond time-scale, is nowadays possible. Our results demonstrate spectral interferometry as a suitable technique for performing a complete spatiotemporal and spatiospectral characterization of such ultrafast beams, whose polarization changes in time and space. The use of spectral interferometry is advantageous as the detection is fully linear (except for the reference measurement) and the data processing is fast, direct and univocal, as well as the acquired data being minimal for this level of measurements. The use of a birefringent plate and a fibre-optic coupler to implement a twofold interferometer avoids using multiple standard interferometers and alignment of beam recombination.

Regarding the vector beam shaping, we show that the combination of spatially varying polarization with temporal polarization shaping can produce singular spatiotemporal polarization dependences. By using a radially polarized beam followed by a narrow polarization gate, we create a complex vector beam with different orientations of the polarization gate or constant circular polarization, depending on the azimuthal coordinate $\theta$ in the transverse plane. We show experimentally that the temporal evolution of the focus of such a beam presents rich dynamics in contrast to focused radially polarized beams. The typical ring mode of focusing is effectively shortened in time because of the narrow polarization gate, which could be used for example to manipulate or trap nanoparticles during shorter times. This can be advantageous when using few-cycle pulses, due to the dispersion in media for ultrabroadband pulses.

In conclusion, the technique presented here constitutes the full spatiotemporally resolved polarization measurements at the femtosecond scale. It can be applied for far-field and near-field measurements, allowing in situ diagnostics within the region of interest in most experiments. As a consequence, a complete characterization of ultrashort vector beams can now be performed, allowing, for example, for quantitative measurements of imperfections and small deviations of spatiotemporal polarization distributions from ideal vector beams (as shown in this work). Such accurate measurements can be of capital relevance in applications in particle trapping, extreme ultraviolet pulse generation through high harmonic generation, or particle acceleration, where the quality of the beam mode is fundamental.

Our results pave the way for the full characterization of the most complex waves created up to now. Some examples include ultrafast beams carrying both spin and orbital angular momentum, whose quality is very important if applied to high nonlinear processes such as high harmonic generation[9], light beams with time-dependent orbital angular momentum properties[10], synthetic chiral fields[14], or bicircular fields composed of counter rotating fields of different frequencies[19,64,65]. The emergence of these complex beams has opened opportunities, for example, to resolve the enantiomeric response of chiral molecules or to perform ultrafast studies of magnetic materials. Further advances in these new directions of ultrafast science are ultimately related to our ability to develop characterization techniques of these new laser sources.

**Methods**

**Experimental details**. The laser beam employed to perform the measurements was delivered by a chirped pulse amplification Ti:sapphire laser system (Spectra Physics Inc.), with central wavelength at 797 nm and spectral bandwidth of 10 nm FWHM (full width at half maximum) at a repetition rate of 1 kHz. When compressed, the laser pulses have a duration of 100 fs FWHM.

The fibre-coupler is made of broadband single-mode fibres centred at 800 nm, being both input arms almost equal-length so that their dispersion is compensated. The relative dispersion due to small difference (~1 mm) between both input arms is calibrated with spectral interferometry.

The thick birefringent plate used was a 3-mm calcite plate (Altechna) with the fast axis oriented vertically. The multiple-order waveplate QWM is a 3-mm quartz plate operating as quarter-wave for 806 nm. The zero-order waveplate QW0 is a 1.3-mm quartz plate designed for quarter-wave operation at 800 nm.

The spectra were acquired with a fibre-coupled spectrometer (Avantes). The spatial scan in the *xy* plane was done with a two-axes motorized stage (Thorlabs). The reference pulse was characterized with a home-made SPIDER measurement[56].

The s-waveplate for 800 nm wavelength was fabricated by ultrafast laser nanostructuring of silica glass.

**Spectral interferometry and data analysis**. In spectral interferometry, two delayed collinear pulses interfere in the spectral domain and their spectral fringes encode their relative phase as given in

$$S_{total}(\omega) = S_1(\omega) + S_2(\omega) + 2\sqrt{S_1(\omega)S_2(\omega)}cos(\phi_1(\omega) - \phi_2(\omega) - \omega\tau) \qquad (2)$$

The relative phase can be extracted by Fourier-transform Spectral Interferometry (FTSI) analysis of the fringes[57], which is detailed in Supplementary Note 2. The pulse delay $\tau$ must be high enough to separate the signals in the time domain after Fourier-transform, as well as small enough to yield spectral interferences within the spectrometer resolution. Note that we perform a twofold interferometer. On one hand, the delay introduced by the birefringent plate between the *x*- and *y*- component of the unknown beam is 1.8 ps, determined by the plate thickness and material (3-mm calcite). In the spectral interferometry between the reference pulse and the *x*-component of the beam, we manually introduced a 2 ps delay. The spectral bandwidth of the unknown pulse must be less or equal than the reference spectrum that interfere, and their relative amplitude must be of the same order to obtain better contrasted fringes. Note that we subtract the individual spectra $S_1(\omega)$ and $S_2(\omega)$ before applying the FTSI algorithm to improve the reconstruction for lower contrasts (details in Supplementary Note 2). The spatial resolution of the technique is related to the mode-field diameter of the single-mode optical fibre, being in the present case 4 μm. The resolution of the motorized 2D translation stage is 1 μm.

As the reference pulse is characterized —and hence, the phase $\phi_{\text{Ref}}(\omega)$ is known—, the phase of the horizontal projection of the unknown vector beam, $\phi_x(\omega; x, y)$, is obtained from the spatially-resolved spectral interferences with the reference pulse, given by

$$S_{x+\text{Ref}}(\omega; x, y) = S_x(\omega; x, y) + S_{\text{Ref}}(\omega) + \\ 2\sqrt{S_x(\omega; x, y)S_{\text{Ref}}(\omega)}cos(\phi_x(\omega; x, y) - \phi_{\text{Ref}}(\omega) - \omega\tau) \tag{3}$$

Then, from the interference between the projections at 45° of the $x$ and $y$ components of the unknown beam, the phase of the vertical projection of the unknown vector beam, $\phi_Y(\omega; x, y)$, is obtained as

$$S_{x+y}(\omega; x, y) = \frac{1}{2}S_x(\omega; x, y) + \frac{1}{2}S_y(\omega; x, y) + \\ \sqrt{S_x(\omega; x, y)S_y(\omega; x, y)}cos\left(\phi_y(\omega; x, y) - \phi_x(\omega; x, y) + \phi_{yx,\text{BP}}(\omega; x, y)\right) \tag{4}$$

where $\phi_{yx,\text{BP}}$ is the relative phase of the birefringent plate eigenaxes. The calibration of this phase is described in the following subsection. Notice that this relative phase is directly responsible for the introduction of the delay $\tau_{\text{YX}}$ used for the interferometry.

In general, the delay $\tau$ introduced by a birefringent plate with thickness $d$ and with extraordinary and ordinary refractive indices $n_e$ and $n_o$, respectively, can be calculated as

$$\tau = \frac{\text{d}}{\text{d}\omega}[\phi_o(\omega) - \phi_e(\omega)] = \frac{\text{d}}{\text{d}\omega}\left[\left(n_o(\omega) - n_e(\omega)\right)\frac{\omega}{c}d\right] \approx (n_o - n_e)\frac{d}{c} \tag{5}$$

where $c$ is the speed of light and we have assumed that the refractive indices are not dispersive. The expression in Eq. (5) is applied to the thick birefringent plate used to introduce $\tau_{\text{YX}}$ for interferometry, as well as for the QWM introducing $\tau_{\text{PG}}$ in the polarization gate.

The spatial scans of the individual spectra, $S_x$ and $S_y$, are optional as they can be obtained with the FTSI algorithm from the measurement of $S_{\text{Ref}}$ and the spectral interferometry scans given above, i.e., the spatially-resolved interferences $S_{x+\text{Ref}}$ and $S_{x+y}$ of Eq. (3) and (4), respectively. Nevertheless, as they can be directly measured, we acquired them, since the performance of the FTSI algorithm is improved when subtracting the individual spectra from the interferences before data processing.

**Calibrations**. The global dispersion of the birefringent plate can be calculated from the thickness and the refractive index using Sellmeier equations. However, the accurate knowledge of the relative phase between the fast and slow axes is critical for the correct retrieval of the beam polarization. This calibration depends on the thickness and alignment of the birefringent plate. In order to calibrate the system we used a linearly polarized pulse, with no time evolving polarization, at 45° before the birefringent plate[53]. From its own interferences, with the projection $S_{x+y}$, we retrieved accurately the relative dispersion of the birefringent plate. In our measurements, we repeated this calibration after any realignment. We also found that the calibration of the birefringent plate did not depend on the transverse spatial position. Furthermore, with the same calibrating pulse (linearly polarized at 45°), we measured $S_x$ and $S_y$ at the same sampling position, to calibrate the amplitude response of the system, which we used to correct the measurements of the individual spectra, $S_x$ and $S_y$, of the unknown beam.

**Models for the simulations**. To simulate the shaped vector beam shown in this work, we started from a homogeneous beam, plane wave, using the experimental spectral amplitude of the laser output. To model the zero-order and multiple-order quartz waveplates, we firstly calibrated their thickness and retardation (if previously unknown) using spectral interferometry in combination with our detection (birefringent plate, linear polarizer and spectrometer). In the simulations, we calculated the dispersion of their eigenaxes from Sellmeier equations and then we imposed the known retardation for the corresponding operation wavelength. Naturally, we applied every element considering the described orientations.

For the s-waveplate, we modelled it as a half-wave plate with the fast axis orientation depending on the azimuthal coordinate $\theta$. As a reference, when the s-waveplate is oriented to create RP from input horizontal linear polarization, the fast axis orientation is $\theta/2$. We operated in the space-frequency domain, and at the end we obtained the space-time dependence by Fourier transformation.

**Generation of the narrow polarization gate**. The experimental implementation for the narrow polarization gate[61] consists of using two consecutive quarter-waveplates, the first multiple-order QWM and the second zero-order QW0, with relative eigenaxes at 45°. In the scheme of Fig. 6, the fast axes of QWM and QW0 are located at 0° and 45°, respectively. To operate as a narrow gate, the system is illuminated with LP pulses at 45° with respect to QWM (upper drawing). The thickness of the QWM introduces a delay (of the order of the pulse duration) between the $x$ (fast) and $y$ (slow) polarization projections. After the QWM, there is a superposition of both delayed pulse projections, producing LP at 0° and LP at 90° at the leading and trailing edges of the pulse, respectively. Due to the dephase of the QWM, there is CP in the centre of the pulse (same amplitude of the projections). When this pulse impinges on the second waveplate (QW0), the leading and trailing edges of the pulse switch to left- and right-handed CP, respectively, while the CP is converted to LP, producing a sharp minimum of ellipticity in the centre of the pulse, known as narrow polarization gate[53,61–63]. Contrarily, if the input light before the waveplates has LP at 0° (lower scheme in Fig. 6), this is preserved after QWM (having a neutral axis at 0°). The LP is converted to constant CP after QW0 because the 45°-angle between the LP pulse and the eigenaxes of the waveplate. For other input LP pulses, the narrow gate will operate differently, as we show and discuss in the corresponding Results Section when the input beam is radially polarized.

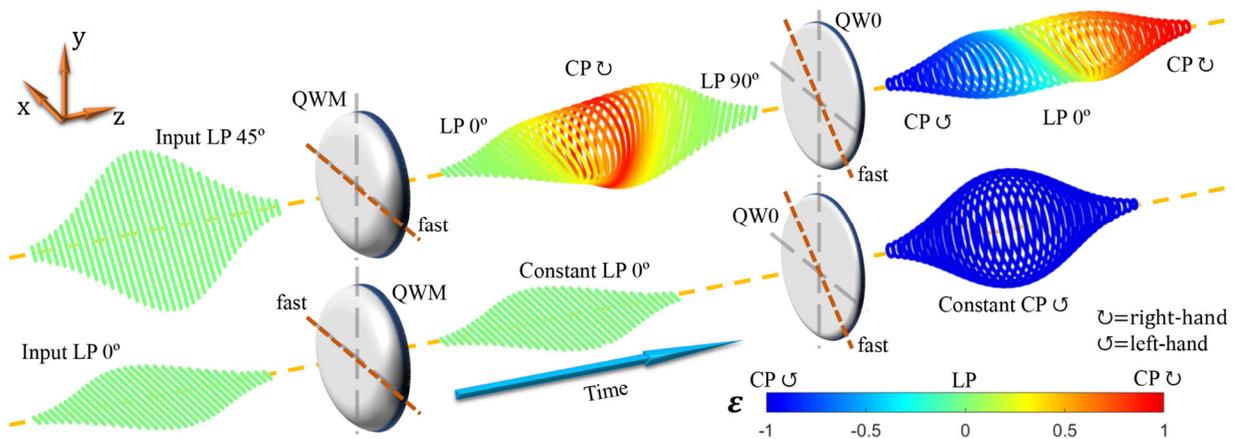

**Figure 6 | Scheme of the operation of the narrow polarization gate**. Firstly, a multiple-order quarter-waveplate (QWM) and, secondly, a zero-order quarter-waveplate (QW0) are oriented with their fast axes at 0° and 45°, respectively. When the input pulse has linear polarization (LP) at 45° (upper drawing), the narrow gate operates normally and produces a pulse with circular polarization (CP) at the edges and LP at the centre. If the input pulse has

LP at 0° (lower drawing), the output pulse presents constant CP. The ellipticity $\varepsilon$ is represented in the given colour scale.

## Data availability

The raw data that support the findings of this article are available from the corresponding author upon reasonable request.

## Acknowledgements

We acknowledge funding from Junta de Castilla y León (SA287P18) and FEDER Funds, and from Spanish Ministerio de Economía y Competitividad (FIS2016-75652-P, FIS2017-87970-R, EQC2018-004117-P). C.H.-G. acknowledges support Ministerio de Ciencia, Innovación y Universidades for a Ramón y Cajal contract (RYC-2017-22745), cofunded by the European Social Fund. B.A. acknowledges funding from the European Union's Horizon 2020 research and innovation programme under the Marie Sklodowska-Curie Individual Fellowship grant agreement No. 798264. PGK acknowledges support of ERC project ENIGMA. This project has received funding from the European Research Council (ERC) under the European Union's Horizon 2020 research and innovation programme (grant agreement No. 851201).

## Author contributions statement

IJS and BA conceived the idea and designed the experiments. BA conducted the experiments. IJS and ILQ participated in the experiments. WH collaborated in the experimental setup. PGK and RD fabricated and provided the s-waveplate. BA analysed the experimental results and performed the simulations and the graphics. IJS, BA and CHG discussed the results and wrote the first manuscript. All authors revised the manuscript.

## Competing interests

The authors declare no competing interests.

# Complete spatiotemporal and polarization characterization of ultrafast vector beams

**Benjamín Alonso, Ignacio Lopez-Quintas, Warein Holgado, Rokas Drevinskas, Peter G. Kazansky, Carlos Hernández-García and Íñigo J. Sola**

**Supplementary Note 1. Definition and parameters of the polarization ellipse**

The electric field in the electromagnetic wave conforming the laser beam is a vector field oscillating as a function of the time. In the paraxial approximation, for a beam propagating in the $z$-axis, the electric field oscillates in the transverse plane (the $xy$ plane). For this initial analysis, we consider a harmonic plane wave, although in the present work we will deal with ultrafast vector beams having frequency-dependent (and thus time-dependent) evolution. We also assume that the light is polarized (i.e. the degree of polarization is maximum), which excludes from the study partially polarized or unpolarized light. The trajectory described by the electric field $\mathbf{E} = \left( E_x, E_y \right)$ in the $xy$ plane is known as the polarization state of the beam, being $E_x$ and $E_y$ the $x$- and $y$-components of the electric field. Those trajectories result from the 2D combination of $E_x$ and $E_y$ during their oscillation at the corresponding light frequency. In the general case, the defined trajectories of $\mathbf{E}$ will be ellipses (see Supplementary Figure 1).

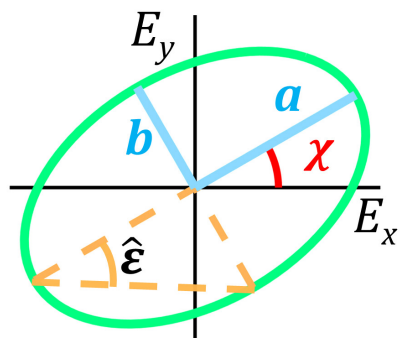

**Supplementary Figure 1 | Polarization ellipse**. Trajectory described by the electric field vector during a period (green), in the general case being an ellipse. The ellipticity is defined as $\varepsilon' = b/a$ and the orientation of the ellipse is determined by the azimuthal angle $\chi$. The ellipticity angle $\hat{\varepsilon}$ is calculated as $\tan|\hat{\varepsilon}| = b/a$.

For the monochromatic (continuous) plane wave in the observation plane (fixed propagation distance $z$), the polarization state of the electric field is determined by the amplitude and phase of the $x$- and $y$- components of $\mathbf{E}$, which are $A_x$, $A_y$, $\phi_x$, and $\phi_y$, respectively. it can be demonstrated that the electric field components satisfy the following expression (Supplementary Reference 1):

$$\left(\frac{E_x}{A_x}\right)^2 + \left(\frac{E_y}{A_y}\right)^2 - 2\frac{E_x}{A_x}\frac{E_y}{A_y}\cos\delta = \sin^2\delta \ , \tag{S1}$$

where the dephase is defined as $\delta = \phi_y - \phi_x$. Actually, the polarization state depends on this relative phase and not on the individual phases. The expression of Eq. (S1) describes a conic, which in this case can be demonstrated to be an ellipse. The ellipse (Supplementary Figure 1) is determined by the ellipticity $\varepsilon' = b/a$ (ratio between the minor axis and the major axis) and the azimuthal angle $\chi$ (the orientation of the major axis with respect to the $x$-axis). The ellipticity angle $\hat{\varepsilon}$ is defined as $\tan|\hat{\varepsilon}| = b/a$ and can be used alternatively to the ellipticity $\varepsilon'$.

The polarization ellipse can degenerate into a line or into a circumference. If the dephase is $\delta = 0$ or $\pi$ ($\cos\delta = 0$), then $E_x$ and $E_y$ oscillate in phase or in opposed phase, and their combination describes a straight line, thus receiving the name of linear polarization. The orientation of this linear polarization is given by $\tan\chi = A_y/A_x$. If the dephase is $\delta = \pm\pi/2$ ($\sin\delta = 0$) and the amplitudes are equal $A_x = A_y$, then $\mathbf{E}$ describes a circumference, which is called circular polarization. In any other case, the light is elliptically polarized.

As said before, the polarization parameters can be determined from $A_x$, $A_y$ and $\delta$. For this, we define auxiliary angle $\alpha$ by $\tan\alpha = A_y/A_x$. Then, it can be shown that $\hat{\varepsilon}$ and $\chi$ can be calculated from (see Supplementary Reference 1) (note that $\varepsilon = \tan\hat{\varepsilon}$):

$$\tan(2\chi) = \tan(2\alpha)\cos\delta \tag{S2}$$

$$\sin(2\hat{\varepsilon}) = \sin(2\alpha)\sin\delta \tag{S3}$$

The handedness of the polarization is given by the sign of $\delta$. The values $\delta > 0$ and $\delta < 0$ describe right- and left-handed polarization, respectively. In order to have a the polarization ellipse fully determined by the parameters $\varepsilon$, $\chi$, we redefine the ellipticity to include the handedness as $\varepsilon = \text{sign}(\delta)\cdot b/a$, so that $\varepsilon = -1$ represents left-handed circular polarization, $\varepsilon = 0$ represents linear polarization, $\varepsilon = +1$ represents right-handed circular polarization, and the values of $0 < |\varepsilon| < 1$ represent elliptically polarized light.

Notice that the previous description stands for harmonic plane waves. In the present work, we explore ultrafast vector beams with complex space and time shaping of the polarization. From the measurements and simulations, we retrieve the full information of the polarization ellipse, described by the parameters $\varepsilon$, $\chi$ and $\delta$ (notice that the information of $\delta$ is complementary for visualization and interpretation, as it can be extracted from $\varepsilon$, $\chi$), together with the local intensity of the beam. Those parameters are calculated for every frequency (and equivalently for every instant of time), and for every position in the $xy$ plane.

**Supplementary Note 2. Extraction of the phase from spectral interferometry**

In this section we explain the general implementation of spectral interferometry and the algorithm used to extract the phase. We consider two delayed collinear pulses that interfere in the spectral domain, being the signal detected with a spectrometer. The resulting spectrum is the sum of the individual spectra of each pulse, $S_1(\omega)$ and $S_2(\omega)$, plus an interference term (crossed term) that encodes the relative spectral phase of the pulses, i.e., $\phi_1(\omega) - \phi_2(\omega)$, as given in the expression

$$S_{\text{total}}(\omega) = S_1(\omega) + S_2(\omega) + 2\sqrt{S_1(\omega)S_2(\omega)}\cos\left(\phi_1(\omega) - \phi_2(\omega) - \omega\tau\right) \qquad (S4)$$

The introduction of the delay $\tau$ between the replicas results in spectral interferences (fringes), which allow us to use the Fourier-transform Spectral Interferometry (FTSI) (Supplementary Reference 2) algorithm to extract the encoded phase difference. The choice of the delay must be appropriate, a too high delay reduces the fringe period so the resolution of the spectrometer may be compromised. On the other hand, the delay must be several times greater than the pulse duration in order to the FTSI algorithm can operate (the signal from the interference term and the individual spectra can be separated).

The relative phase, $\phi_1(\omega) - \phi_2(\omega)$, is encoded in the cosine term. The inversion of the cosine presents ambiguity, which would prevent the direct extraction of the relative phase. For this reason, the fringe analysis with the FTSI algorithm is suitable (see the scheme in Supplementary Figure 2). In a first step, the interference spectrum $S_{\text{total}}(\omega)$ (Supplementary Figure 2a) is inverse Fourier-transformed, resulting in three peaks in the temporal domain (Supplementary Figure 2b). The central peak at $t = 0$ stands for the signal of the individual spectra of each pulse, $S_1(\omega)$ and $S_2(\omega)$. The two lateral peaks, centred at $t = +\tau$ and $t = -\tau$, result from the cosine term being unfolded with the Euler's formula $2\cos\beta = e^{+i\beta} + e^{-i\beta}$, thus corresponding to conjugated terms. By isolating any of these two terms, the relative phase can be extracted. To do this, we choose to filter the peak at $t = +\tau$ with a numerical gate (Supplementary Figure 2c). The filtered peak in the temporal domain is Fourier-transformed to the spectral domain, where the resulting signal corresponds to

$$S_{t=+\tau}(\omega) = \sqrt{S_1(\omega)S_2(\omega)}\exp\left[\mathrm{i}\left(\phi_1(\omega) - \phi_2(\omega) - \omega\tau\right)\right] \qquad (S5)$$

From this expression, the phase difference can be extracted from the argument of the complex magnitude, thus being obtained without ambiguity. The delay can be calculated from the position of the lateral peak $t = +\tau$.

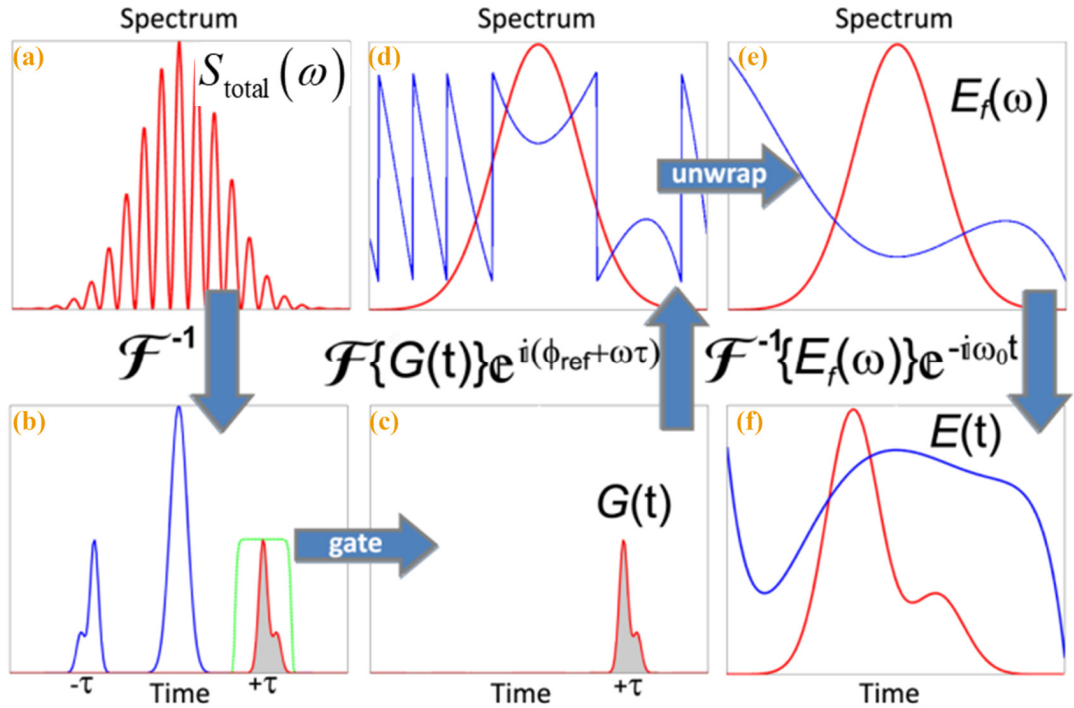

**Supplementary Figure 2 | Scheme of the algorithm implemented for Fourier-transform Spectral Interferometry (FTSI).** First, an inverse Fourier-transform ($\mathcal{F}^{-1}$) is applied to the spectral interferences (a). In the time domain (b), one lateral peak is gated and direct Fourier-transformed ($\mathcal{F}$) (c). In frequency domain, the term $\omega\tau$ and the reference phase are corrected (d), and the phase is unwrapped (e). This gives the spectral amplitude and phase of the test pulse (e), which can be translated to the temporal domain by applying again an inverse Fourier-transform (f). In our case, we measure the spectral amplitude of the pulse directly with the spectrometer. In panels (d-f) the amplitudes and phases are coloured red and blue, respectively. In panel (f) the phase is depicted with the carrier contribution suppressed. Figure extracted from Supplementary Reference 3.

The advantage of using spectral interferometry is that when a pulse is known, for example the pulse 2, it can act as a reference to characterize the pulse 1 (supposed to be unknown). Indeed, if the phase $\phi_2(\omega)$ is known, from Equation S5 we can obtain the phase of the unknown pulse, $\phi_1(\omega)$

. If we further measure the spectrum of the unknown pulse, we can calculate the spectral amplitude $A_1(\omega) = \sqrt{S_1(\omega)}$. Therefore, we have the full information of the pulse in the spectral domain, $E_1(\omega) = A_1(\omega)\exp\left[i\phi_1(\omega)\right]$ (Supplementary Figure 2d). The spectral phase is defined between $-\pi$ and $+\pi$, so it may be unwrapped (this is not a requirement) for convenience of the representation or analysis (Supplementary Figure 2e). By inverse Fourier-transform ($\mathcal{F}^{-1}$), the electric field in the temporal domain can be calculated (Supplementary Figure 2f).

In the experiments, in order to improve the calculation of the relative phase, the individual spectra of each pulse, $S_1(\omega)$ and $S_2(\omega)$, can be subtracted to the total spectrum given in Eq. (S5), obtaining the interference spectrum given in Eq. (S6). By doing so, the central peak in Supplementary Figure 2b is depleted and the lateral peak is more easily isolated. We follow this approach in the present work.

$$S_{\text{interf}}(\omega) = 2\sqrt{S_1(\omega)S_2(\omega)}\cos\left(\phi_1(\omega) - \phi_2(\omega) - \omega\tau\right) \qquad (S6)$$

As detailed in the manuscript, in the present work we use a known reference pulse (created with the beam splitter) to perform spatially-resolved spectral interferometry with the $x$-projection of the unknown vector beam. The unknown beam and the reference pulse are collected and combined with the fibre-optic coupler. The longitudinal position of the collection fibre input of the reference pulse is used to adjust a suitable delay of 2 ps (between the reference pulse and the unknown beam). The reference pulse is not spatially scanned, so its phase is constant and equal to $\phi_{\text{Ref}}(\omega)$, whereas the unknown beam is scanned in the $xy$ plane, therefore obtaining $\phi_x(\omega; x, y)$ from the FTSI algorithm.

Then, the interference between the $x$- and $y$-projections of the unknown beam is measured. For this, the birefringent plate introduces a fixed delay between both components. The linear polarizer set at 45° selects an intermediate projection, in order to exhibit interferences. The spectral interference is scanned in the $xy$ plane. From the FTSI algorithm, the relative phase is obtained. Since the phase $\phi_x(\omega; x, y)$ was characterized before, then the phase $\phi_y(\omega; x, y)$ can be reconstructed. As explained in the Methods section in the manuscript, the spectral amplitudes are obtained from the experimental spatially-resolved spectra of the $x$- and $y$-projections, thus having the full characterization of the beam. By inverse Fourier-transforming that signal, it is obtained the spatiotemporal polarization evolution of the ultrafast vector beam.

**Supplementary Note 3. Characterization of the collimated input laser beam**

In this Section we show the measurement of the input laser beam used for the experiments. As in the first part of the manuscript, we perform a circular scan with radius $R$=3 mm (corresponding to the half-maximum of the beam intensity). In the Supplementary Figure 3a we show the spatially-resolved spectrum as a function of the azimuthal coordinate $\theta$ and in the Supplementary Figure 3b the corresponding spatially-resolved temporal intensity. If the input laser had a homogeneous profile or a perfect Gaussian profile, the spectrum and intensity would be constant for every $\theta$. The actual spatial profile of the beam is shown in Supplementary Figure 3c, where the beam is not symmetric and has inhomogeneities. The circular scan performed is superimposed in Supplementary Figure 3c. The spatial inhomogeneities will be present in the polarization shaped beam presented in the manuscript, as we discuss both for the collimated and the focused measurements.

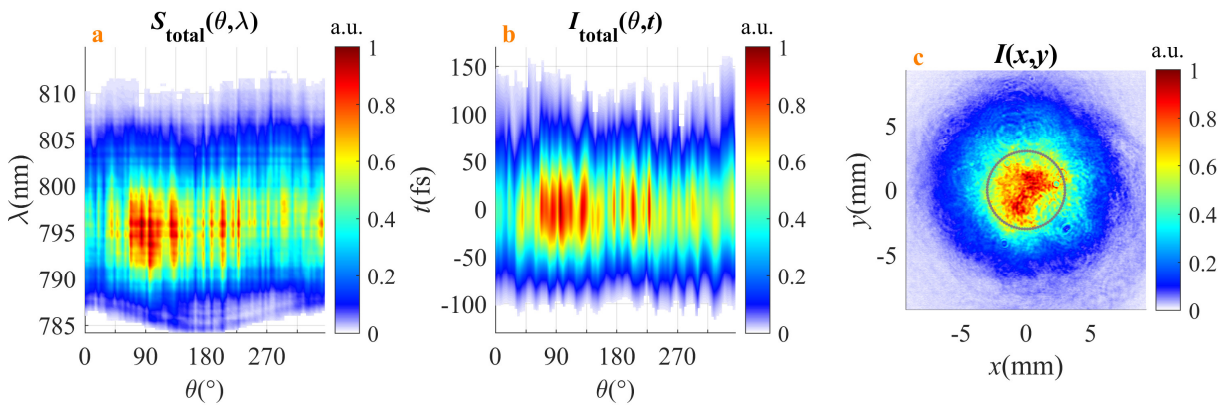

**Supplementary Figure 3 | Measurement of the input laser beam**. **a** Spatially resolved spectrum, and **b** spatiotemporal intensity. **c** Spatial profile of the input beam with the circular scan with radius $R$=3 mm indicated by the 90 scanning points in grey.

**Supplementary Note 4. Characterization and simulation of radially polarized laser pulses**

In this Section we show the experimental measurements compared to the theoretical simulations of a radially polarized beam (Supplementary Figure 4 and Supplementary Figure 5), corresponding to the manuscript Results subsection Characterization of radially polarized laser pulses.

As discussed in the manuscript, the results are achromatic, so ideally the spectrum/temporal intensities, and the polarization parameters are constant in frequency (equivalently wavelength $\lambda$) and time.

For these measurements we perform the circular scan in the azimuthal coordinate $\theta$ described in the previous section.

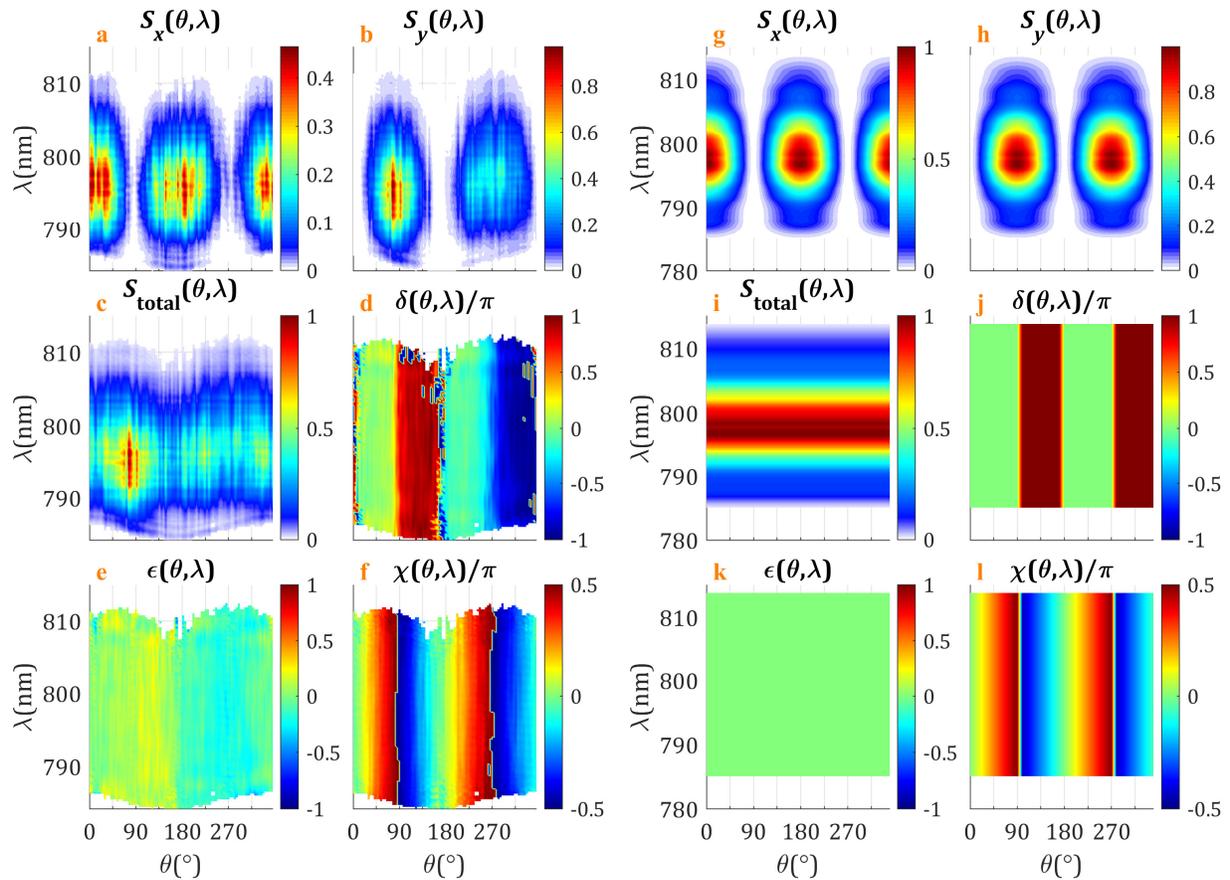

**Supplementary Figure 4 | Measurement (a-f) and simulation (g-l) of the radially polarized beam in the spatiospectral domain**. **a,g** horizontal projection of the spectrum, **b,h** vertical projection of the spectrum, **c,i** total spectrum, **d,j** dephase, **e,k** ellipticity, **f,l** azimuthal angle. The maximum of the total spectrum is normalized to 1.

As for radial polarization, the $x$-polarization projection presents two lobes centred at the azimuthal coordinate $\theta = 0°$ and $\theta = 180°$, while the $y$-polarization projection presents two lobes centred at the azimuthal coordinate $\theta = 90°$ and $\theta = 270°$.

The total spectrum/intensity is constant in $\theta$. The polarization azimuth $\chi$ corresponds to the radial direction for every $\theta$, as expected, and the ellipticity is 0 as in the radial polarization distribution, the beam is locally linearly polarized.

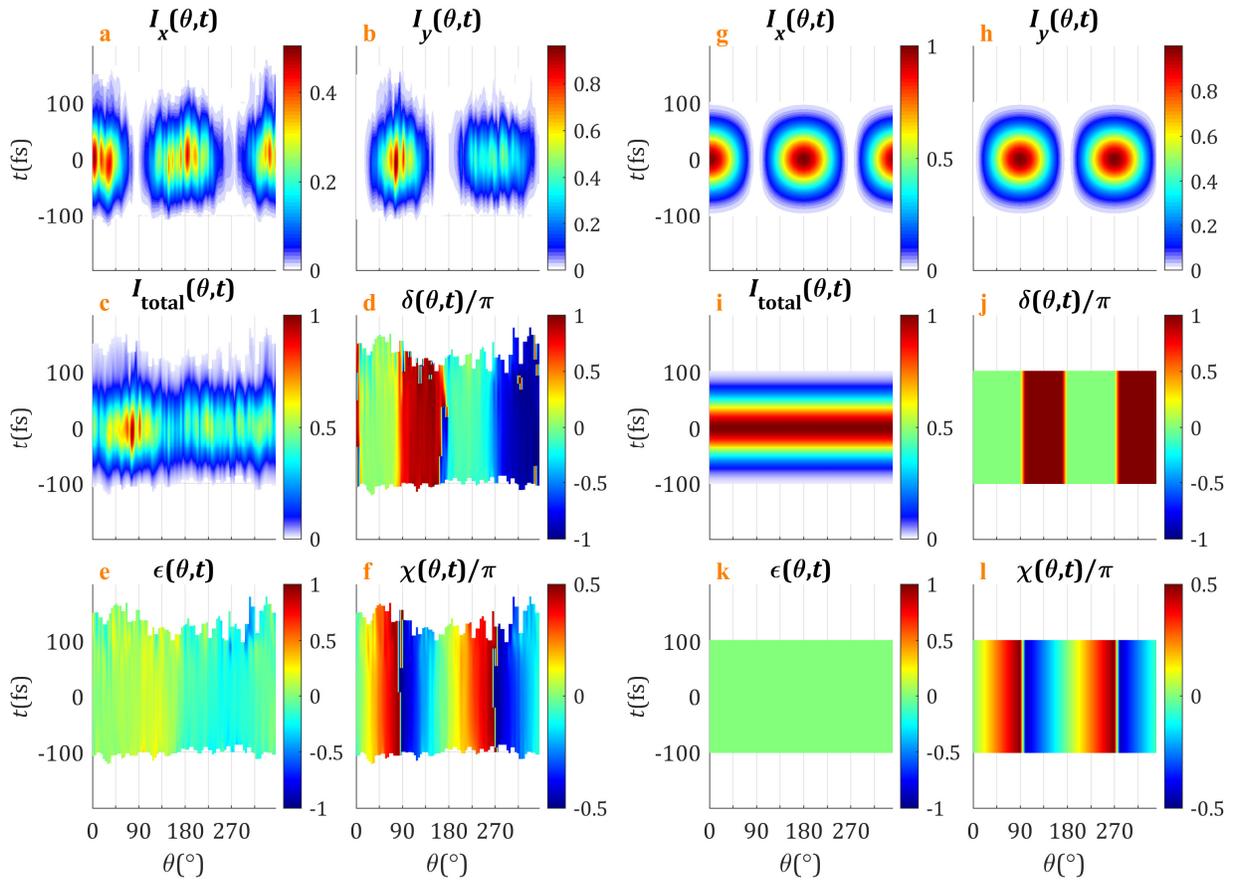

**Supplementary Figure 5 | Measurement (a-f) and simulation (g-l) of the radially polarized beam in the spatiotemporal domain**. **a,g** horizontal projection of the intensity, **b,h** vertical projection of the intensity, **c,i** total intensity, **d,j** dephase, **e,k** ellipticity, **f,l** azimuthal angle. The maximum of the total intensity is normalized to 1.

**Supplementary Note 5. Simulation of azimuthally polarized laser pulses**

As a complementary resource, in this Section we show the simulations of an azimuthally polarized beam (Supplementary Figure 6). The results are similar than for the previous case, with the exception that they are shifted 90° in the azimuthal coordinate $\theta$. The two lobes of the *x*-polarization projection are centred at the azimuthal coordinate $\theta = 90°$ and $\theta = 270°$, while for the *y*-polarization projection they are centred at the azimuthal coordinate $\theta = 0°$ and $\theta = 180°$. The polarization azimuth $\chi$ corresponds to the azimuthal direction for every $\theta$, i.e., it is perpendicular to the radial direction.

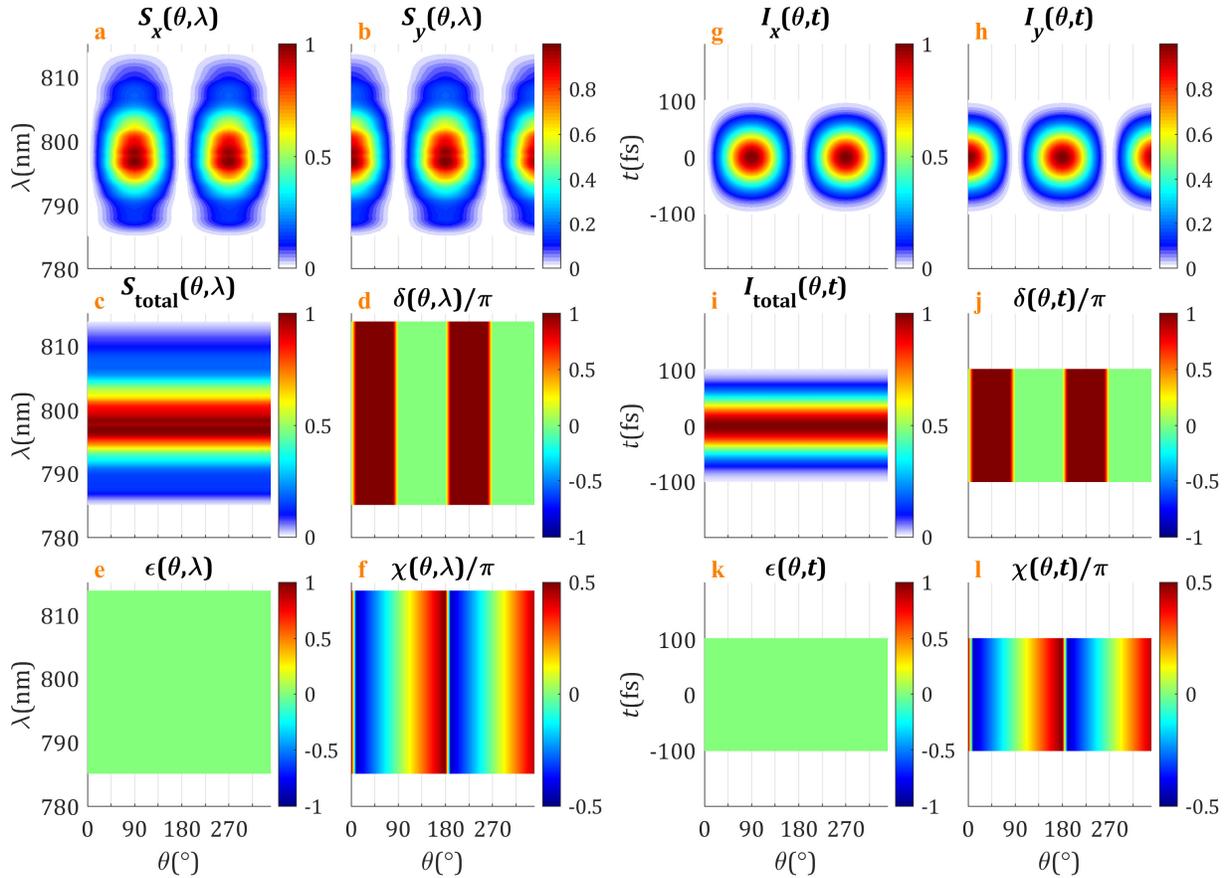

**Supplementary Figure 6 | Spatiospectral (a-f) and spatiotemporal (g-l) domain simulation of the azimuthally polarized beam**. **a,g** horizontal projection of the spectrum/intensity, **b,h** vertical projection of the spectrum/intensity, **c,i** total spectrum/intensity, **d,j** dephase, **e,k** ellipticity, **f,l** azimuthal angle. The maximum of the total spectrum/intensity is normalized to 1.

**Supplementary Note 6. Measuring shaped time-dependent ultrafast vector beams**

In this Section we show the experimental measurements compared to the theoretical simulations of a radially polarized beam followed by a narrow polarization gate (Supplementary Figure 7 and Supplementary Figure 8), corresponding to the manuscript Results subsection Measuring shaped time-dependent ultrafast vector beams. Some of these results have already been presented and explained in the manuscript. Here, the complementary graphics of the results described in the manuscript are presented to help the reader to visualize them.

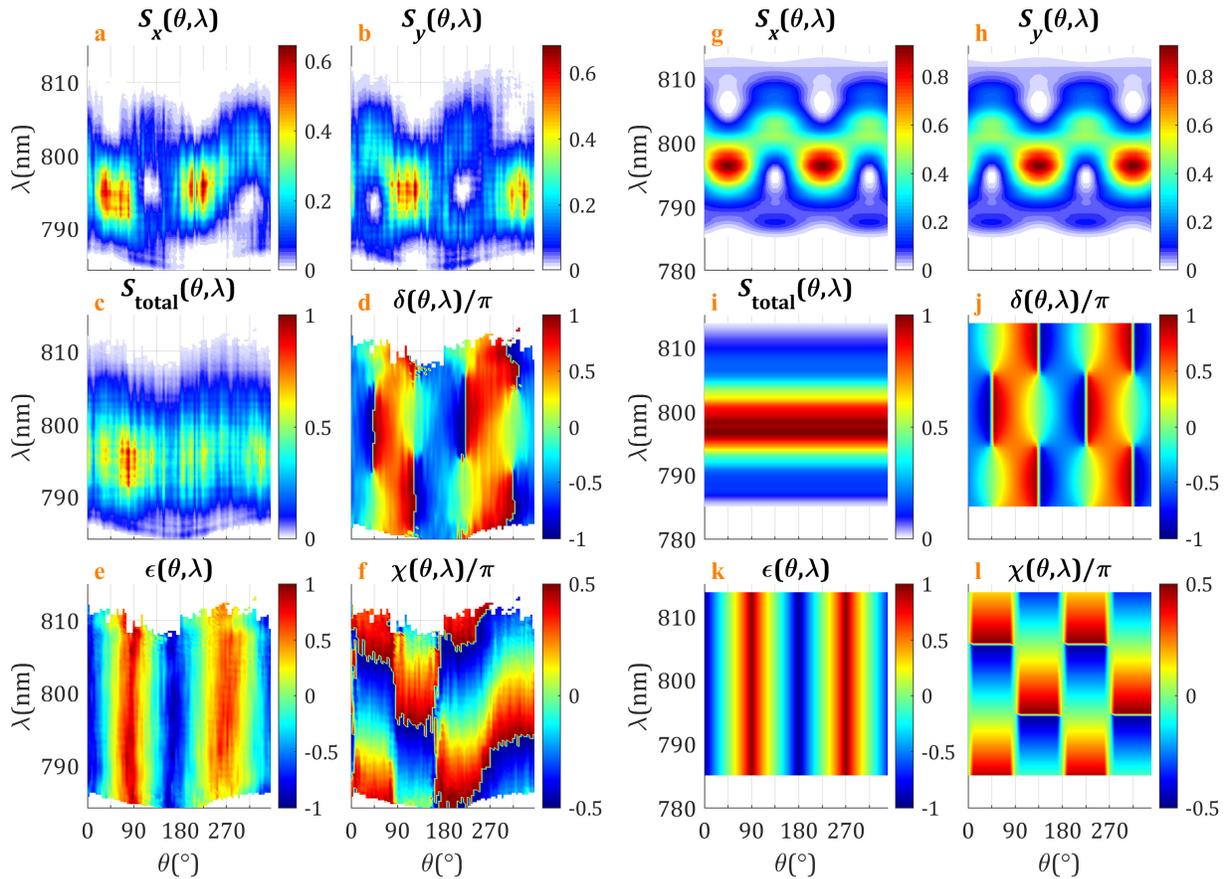

**Supplementary Figure 7 | Measurement (a-f) and simulation (g-l) of the radially polarized beam followed by a narrow polarization gate in the spatiospectral domain. a,g** horizontal projection of the spectrum, **b,h** vertical projection of the spectrum, **c,i** total spectrum, **d,j** dephase, **e,k** ellipticity, **f,l** azimuthal angle. The maximum of the total spectrum is normalized to 1.

For example, in Supplementary Figure 8e and 8k, it is seen how the beam has linear polarization at the times where the left-handed and right-handed circular polarization pulses have the same amplitude for each value of the azimuthal coordinate. In the *x*-projection of the intensity (Supplementary Figure 8a and 8g), the zeros correspond to the positions of the centre of the narrow polarization gate where the beam has vertical linear polarization. Similarly, in the *y*-projection of the intensity (Supplementary Figure 8b and 8h), the zeros correspond to the positions of the centre of the narrow polarization gate where the beam has horizontal linear polarization. It is seen how the centre of the narrow polarization gate is shifted back and forward in time then varying the azimuthal coordinate (which varies the relative amplitude of the left-handed and right-handed circular polarization pulses).

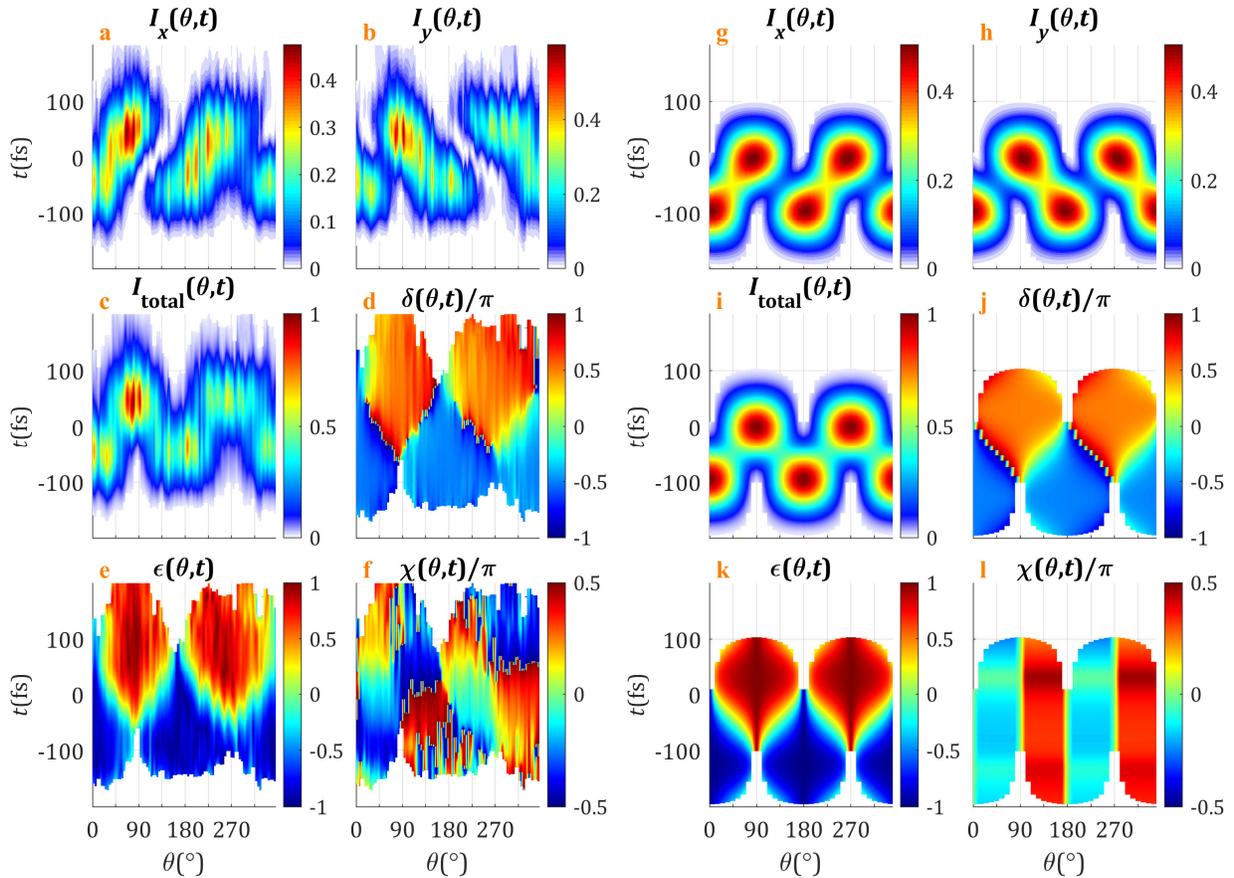

**Supplementary Figure 8 | Measurement (a-f) and simulation (g-l) of the radially polarized beam followed by a narrow polarization gate in the spatiotemporal domain.** **a,g** horizontal projection of the intensity, **b,h** vertical projection of the intensity, **c,i** total intensity, **d,j** dephase, **e,k** ellipticity, **f,l** azimuthal angle. The maximum of the total intensity is normalized to 1.

# Supplementary Note 7. Scanning 2D grid used for the focusing monitorization of time-dependent vector beam pulses

In this Section we show the scanning 2D grid used for the experimental measurements of the focusing of a radially polarized beam and the focusing of a radially polarized beam followed by a narrow polarization gate, corresponding to the manuscript Results subsection Focusing monitorization of time-dependent vector beam pulses. As discussed in the manuscript, the focused beam parameters depend not only in the azimuthal but also in the radial coordinate, so the whole $xy$ plane must be scanned. For this purpose, we performed a 2D square scan in a regular grid of 21×21 points, using a scanning step is 18 μm.

As seen in the spatial profile of Supplementary Figure 9a, the full transverse profile is covered with that grid. In Supplementary Figure 9b, we show how the scanning grid is overlaid with the profile of the beam.

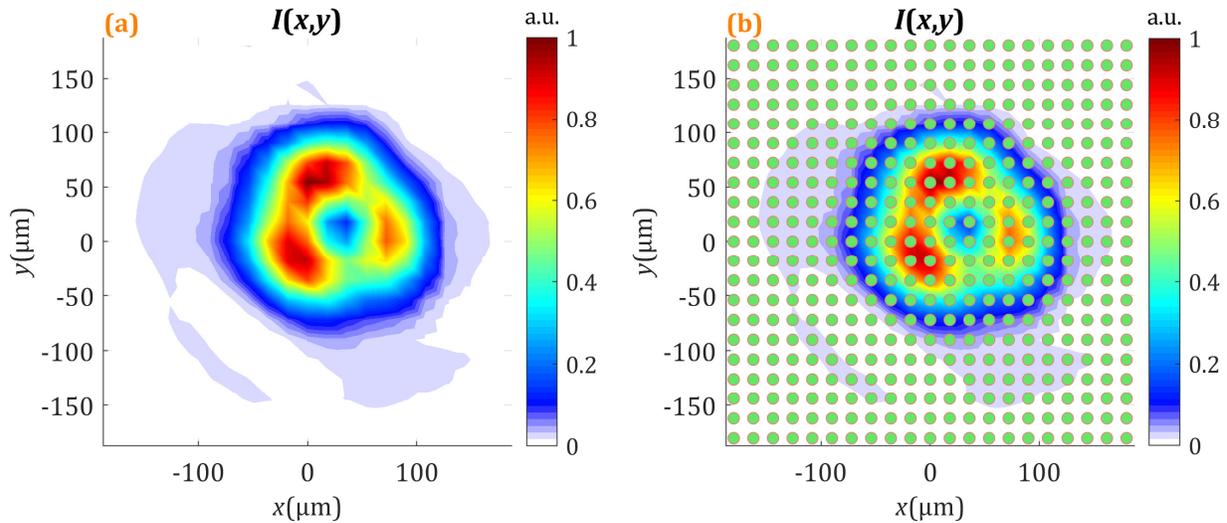

**Supplementary Figure 9 | Spatial profile of the focused beam and scanning grid. a** Spatial profile of the radially polarized focused beam (obtained integrating in frequency the spectrometer signal). **b** The same spatial profile where the scanning grid used to sample the beam has been superimposed. The scanning step is 18 μm in a square grid of 21×21 points.

## Supplementary References